\documentclass{aa}
\usepackage{subfigure}
\usepackage{graphicx}
\usepackage{txfonts}
\usepackage{natbib}
\bibpunct{(}{)}{;}{a}{}{,}

\def \kms{km s$^{-1}$}
\def \cmiii{cm$^{-3}$}

\begin{document}

\title{Detection of 6.7 GHz methanol absorption towards hot corinos}
\author{J. D. Pandian\inst{1} \and S. Leurini\inst{2} \and K. M. Menten\inst{1} \and A. Belloche\inst{1} \and P. F. Goldsmith\inst{3}}
\institute{Max-Planck-Institut f\"{u}r Radioastronomie, Auf dem H\"{u}gel 69, 53121 Bonn, Germany\\
\email{[jpandian;kmenten;belloche]@mpifr-bonn.mpg.de}
\and
ESO, Karl-Schwarzschild Strasse 2, 85748 Garching-bei-M\"{u}nchen, Germany\\
\email{sleurini@eso.org}
\and
Jet Propulsion Laboratory, Pasadena, CA 91109, U.S.A.\\
\email{Paul.F.Goldsmith@jpl.nasa.gov}}

\abstract
{Methanol masers at 6.7 GHz have been found exclusively towards high-mass star forming regions. Recently, some Class 0 protostars have been found to display conditions similar to what are found in hot cores that are associated with massive star formation. These hot corino sources have densities, gas temperatures, and methanol abundances that are adequate for exciting strong 6.7 GHz maser emission.}
{This raises the question of whether 6.7 GHz methanol masers can be found in both hot corinos and massive star forming regions, and if not, whether thermal methanol emission can be detected.}
{We searched for the 6.7 GHz methanol line towards five hot corino sources in the Perseus region using the Arecibo radio telescope. To constrain the excitation conditions of methanol, we observed thermal submillimeter lines of methanol in the NGC1333-IRAS~4 region with the APEX telescope.}
{We did not detect 6.7 GHz emission in any of the sources, but found absorption against the cosmic microwave background in NGC1333-IRAS~4A and NGC1333-IRAS~4B. Using a large velocity gradient analysis, we modeled the excitation of methanol over a wide range of physical parameters, and verify that the 6.7~GHz line is indeed strongly anti-inverted for densities lower than 10$^{6}$~\cmiii.  We used the submillimeter observations of methanol to verify the predictions of our model for IRAS~4A by comparison with other CH$_3$OH transitions. Our results indicate that the methanol observations from the APEX and Arecibo telescopes are consistent with dense (n $\sim 10^6$~\cmiii), cold ($T\sim 15-30$~K) gas.}
{The lack of maser emission in hot corinos and low-mass protostellar objects in general may be due to densities that are much higher than the quenching density in the region where the radiation field is conducive to maser pumping.}


\maketitle

\section{Introduction}
The $5_1-6_0$ A$^+$ line of methanol at 6.7 GHz is the strongest of
Class II methanol masers. 6.7 GHz masers are unique compared to OH and
H$_2$O masers in that they are associated exclusively with high-mass
star forming regions. For example, targeted searches towards low-mass
young stellar objects have not resulted in any detections of 6.7 GHz
methanol masers \citep{mini03,bour05,xu08}. Theoretical models suggest
that the maser emission switches on at gas densities greater than
$\sim 10^4$~\cmiii and methanol fractional abundance greater than $\sim
10^{-7}$, along with the requirement of a far-infrared radiation field
to excite methanol to torsionally excited states as part of the
pumping cycle \citep{sobo97, crag02, crag05}. These conditions are
found in high-mass star forming regions where the far-infrared
radiation field is generated by warm dust ($T_{D} \sim 100-150$ K),
and high methanol abundances are achieved by evaporating the mantles
of dust grains, either by the passage of a shock or through the
radiation field of the massive young stellar object. The molecules
released from the grains further trigger formation of more complex
molecules in warm gas chemistry (e.g. \citealt{char92}).

Recently, there has been evidence of hot core chemistry in low-mass
protostars. For example, complex molecules such as HCOOCH$_3$, HCOOH,
and CH$_3$CN have been found in IRAS 16293--2422 \citep{caza03} and
NGC1333-IRAS~4A \citep{bott04} along with high abundance of H$_2$O,
CH$_3$OH, and H$_2$CO \citep{cecc00a,cecc00b, scho02}. The abundance of
methanol in the compact, warm regions in the immediate proximity of these embedded sources, known as hot corinos, can be as high as $7\times 10^{-7}$
(\citealt{mare05} and references therein). Moreover, the regions have
warm temperature ($> 100$ K) and high density ($> 10^7$ \cmiii;
\citealt{cecc00a}). This raises the question of whether hot corinos
can excite maser emission in Class II methanol lines. A discovery of
Class II methanol masers in hot corinos would force
re-interpretation of the methanol maser data in the literature. In
this paper, we report on the results of a search for 6.7 GHz methanol
masers towards five hot corino sources in the Perseus region.

\section{Observations}
\subsection{Arecibo Observations}
The search for 6.7 GHz maser emission towards hot corinos was carried out using the 305 m Arecibo radio telescope\footnote{The Arecibo Observatory is part of the National Astronomy and Ionosphere Center, which is operated by Cornell University under a cooperative agreement with the National Science Foundation.} in October 2006. The C-High receiver was used along with the ``interim'' correlator backend. Two units of the correlator were used with 1.5625 MHz bandwidth to measure two orthogonal linear polarizations with 9-level sampling and 1024 channels per polarization giving a velocity coverage of 140 \kms~with 0.28 \kms~resolution after Hanning smoothing. The adopted rest frequency for the transition was 6668.518 MHz; a more accurate measurement of 6668.519 MHz by \citet{brec95} gives an error of 0.04 \kms~in the velocity scale, which is well below the resolution of the spectra. The sources observed and their coordinates are indicated in Table 1. Sources NGC1333-IRAS~4A and NGC1333-IRAS~4B had on-source integration times of 45 minutes each, while the remaining sources had on-source integration times of 30 minutes each. The data were taken in position switched mode. The ``off'' spectra were taken over the same elevation and azimuth range as for the ``on'' spectra to ensure cancellation of standing waves. During each observation, the pointing was checked using a continuum source and was found to be accurate within 10\arcsec. The system temperature was between 23 and 34 K, with approximately 20\% difference between the two polarizations. The half-power beamwidth (HPBW) of the telescope at this frequency is $\sim 40$\arcsec.

\begin{table}
\caption{Hot corinos observed at 6.7 GHz. The columns show source name, equatorial J2000 coordinates and the r.m.s. noise in spectra.}
\label{table1}
\centering
\begin{tabular}{lccc}
\hline \hline
Source & $\alpha$ & $\delta$ & $S$ (rms) \\
& (J2000) & (J2000) & mJy \\
\hline
L1448--N.................... & 03 25 36.6 & +30 45 15 & 1.2 \\
L1448--MM................ & 03 25 38.8 & +30 44 05 & 1.3 \\
NGC1333--IRAS2...... & 03 28 55.4 & +31 14 35 & 1.2 \\
NGC1333--IRAS~4A... & 03 29 10.3 & +31 13 31 & 1.0 \\
NGC1333--IRAS~4B... &  03 29 12.0 & +31 13 09 & 1.0 \\
\hline
\end{tabular}
\end{table}

The data were reduced in IDL using procedures maintained by the
observatory. The antenna temperature was derived using noise diodes,
which was converted to flux density using an elevation and azimuth
dependent gain curve for this frequency, the typical conversion factor
being 5 K/Jy. The resulting r.m.s. noise in the spectra after Hanning
smoothing was 1.0 mJy for IRAS~4A and IRAS~4B, and 1.2--1.3 mJy for
the remaining sources. We did not carry out any baselining of the
spectra.

\subsection{APEX Observations}
It is possible to carry out an excitation analysis to reproduce the observed 6.7 GHz feature using millimeter and submillimeter lines of methanol \citep{leur04}. In order to do this, we observed the NGC1333-IRAS~4 region in the $(6_K-5_K)$~CH$_3$OH band at $\sim 290$~GHz using the APEX 12 m telescope\footnote{This publication is based on data acquired with the Atacama Pathfinder Experiment (APEX). APEX is a collaboration between the Max-Planck-Institut f\"{u}r Radioastronomie, the European Southern Observatory, and the Onsala Space Observatory}. The observations were carried out in November 2006 using the APEX2A double side band receiver \citep{risa06} and the Fast Fourier Transform Spectrometer (FFTS; \citealt{klei06}). The receiver was tuned to a frequency of 290.300 GHz in the lower sideband. One unit of the FFTS was centered at the tuning frequency, while the second unit was centered at 291.5 GHz. Both units had a bandwidth of 1 GHz with 16\,384 channels, the effective velocity resolution being 0.13 \kms~(since the noise in successive channels is correlated; \citealt{klei06}). The pointing was checked approximately every 1.5 hours in CO $(J=3-2)$ on IK-Tau and was found to be accurate to within 2\arcsec. The focus was verified using Saturn. The system temperatures were typically 120--150 K in units of $T_A^*$. The half-power beamwidth of the telescope at this frequency is around 21\arcsec. A $1.5\arcmin \times 2.5\arcmin$ region was mapped using the on-the-fly scanning mode with a step size of 6\arcsec.

The data were reduced using the CLASS software. A first order polynomial was fitted to the line-free channels and subtracted off the baseline. The data were converted to main beam temperature units ($T_{MB}$) assuming a beam efficiency of 0.73 and a forward efficiency of 0.97 \citep{gues06}. Individual lines were then identified using the XCLASS software \citep{comi05}.

\section{Results}\label{sec_res}

\begin{figure}
\centering
\includegraphics[width=\columnwidth]{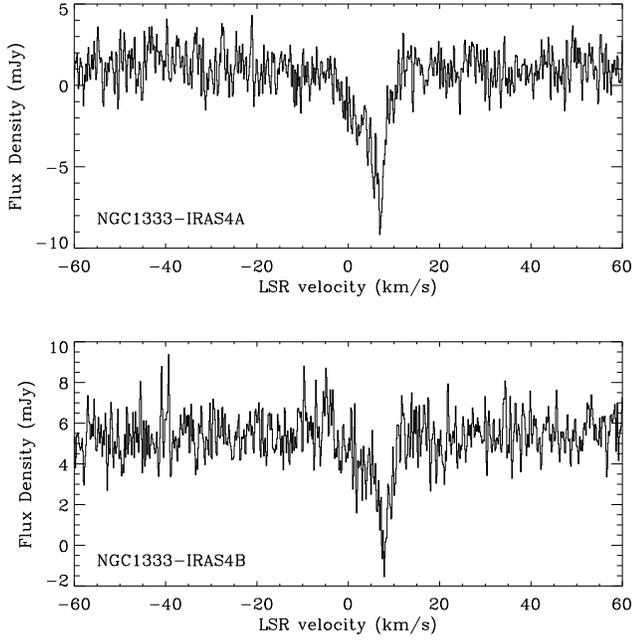}
\caption{The observed 6.7 GHz spectra showing absorption in NGC1333-IRAS~4A and IRAS~4B.}\label{abs67o}
\end{figure}

\begin{figure}
\centering
\includegraphics[width=\columnwidth]{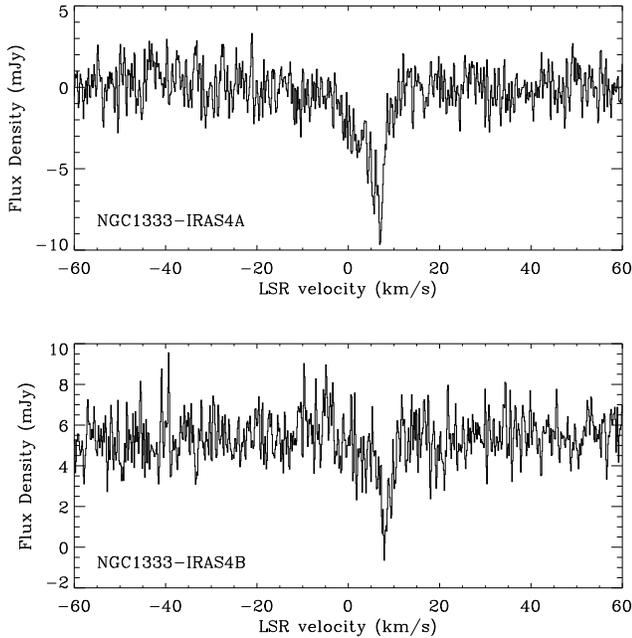}
\caption{6.7 GHz spectra of NGC1333-IRAS~4A and IRAS~4B that have been deconvolved to remove their mutual contamination. A comparison with Fig. \ref{abs67o} reveals that the spectrum of IRAS~4A is almost identical, while IRAS~4B shows a more significant difference.}\label{abs67d}
\end{figure}

We did not detect 6.7 GHz maser emission in any of the sources, but detected prominent absorption in NGC1333-IRAS~4A and NGC1333-IRAS~4B, as shown in Fig. \ref{abs67o}. IRAS~4A and IRAS~4B are separated by 31\arcsec, which is less than the half-power beamwidth of 40\arcsec. Thus, the measured spectrum towards IRAS~4A has some contribution from IRAS~4B and vice versa. Assuming a symmetric beam, if $\alpha$ is the contribution of the second source at the position of the first, then the measured spectra can be written as
\begin{eqnarray}
S^m_{\mathrm{IRAS4A}} & = & S^a_{\mathrm{IRAS4A}} + \alpha S^a_{\mathrm{IRAS4B}} \\
S^m_{\mathrm{IRAS4B}} & = & S^a_{\mathrm{IRAS4B}} + \alpha S^a_{\mathrm{IRAS4A}}
\end{eqnarray}
where $S^a$ and $S^m$ indicate the actual and measured spectra respectively. Since the beam is a Gaussian to good approximation at 31\arcsec~from center, the parameter $\alpha$ is calculated as 0.19. Equations (1) and (2) can then be inverted to calculate the actual spectra, the results being shown in Fig. \ref{abs67d}. The contribution of IRAS~4B to the observed IRAS~4A spectrum is small (less than $\sim$ 10\%), while the contribution of IRAS~4A to the observed IRAS~4B spectrum is much more significant ($\sim$ 30\%).

\begin{figure}
\centering
\includegraphics[width=\columnwidth]{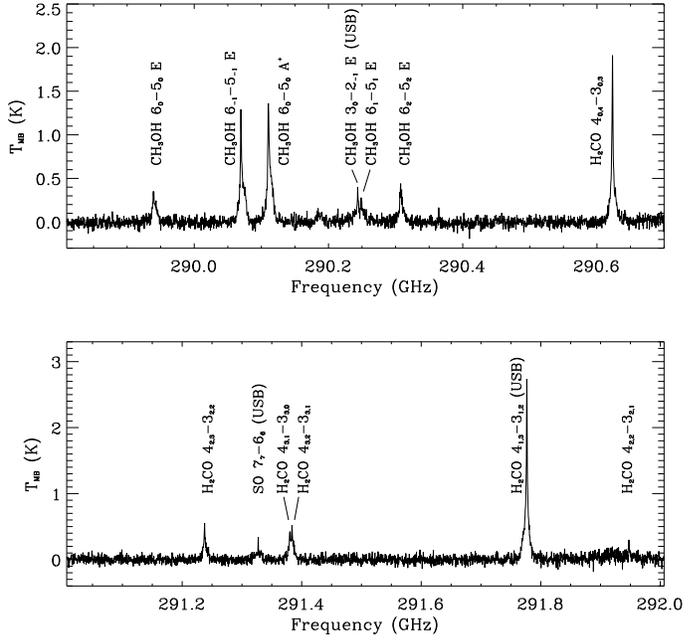}
\caption{APEX spectra of NGC1333-IRAS~4A smoothed to the Arecibo resolution of 40$''$ (HPBW).}\label{apexiras4a}
\end{figure}

\begin{figure}
\centerline{\resizebox{1.0\hsize}{!}{\includegraphics[angle=270]{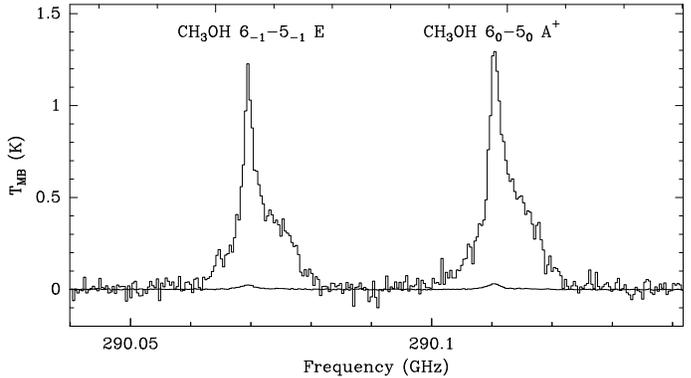}}}
\caption{The higher intensity spectrum shows two 290 GHz methanol lines from the APEX data towards IRAS~4A that is smoothed to 40$''$ resolution and includes contributions from IRAS~4B. The weaker spectrum represents the difference between that spectrum and another 40$''$ resolution spectrum from the same position that does not include data in the direction of IRAS~4B. The ``contamination'' from IRAS~4B is thus estimated to be at most 3.5\%.}\label{4bcontamination}
\end{figure}

For modeling the observed 6.7 GHz absorption, we smoothed the APEX data to the spatial scale of the Arecibo beam. The resulting spectrum at the IRAS~4A position is shown in Fig. \ref{apexiras4a}. We were unable to obtain spectra from the smoothed data cube at the IRAS~4B position since the footprint of the Arecibo beam at this location extends beyond the eastern edge of the APEX map. Since the convolution will give rise to the same degree of blending between the two sources as in the Arecibo observation, one would ideally have to deconvolve the spectra of the two sources extracted from the smoothed data cube, using the same parameters as for the Arecibo data. In the absence of spectra from both IRAS~4A and IRAS~4B at the same spatial resolution, it is not possible to deconvolve the two sources as was carried out for the Arecibo data.
To quantify this contamination in the APEX data, we extracted a spectrum from IRAS~4A taking care to exclude the IRAS~4B region during the spatial convolution, and compared it with the spectrum at the same position from the original smoothed datacube (which includes the contribution from IRAS~4B). Fig \ref{4bcontamination} shows the difference between the two spectra overlaid on the IRAS~4A spectrum from the smoothed datacube. From this analysis, we estimate the contribution of IRAS~4B to IRAS~4A in the APEX data to be at most 3.5\%. Henceforth, we will ignore the contamination of IRAS~4B on IRAS~4A. Since the contribution of IRAS~4A to the IRAS~4B spectrum is significant, we do not model IRAS~4B.

\begin{figure}
\centering
\includegraphics[width=\columnwidth]{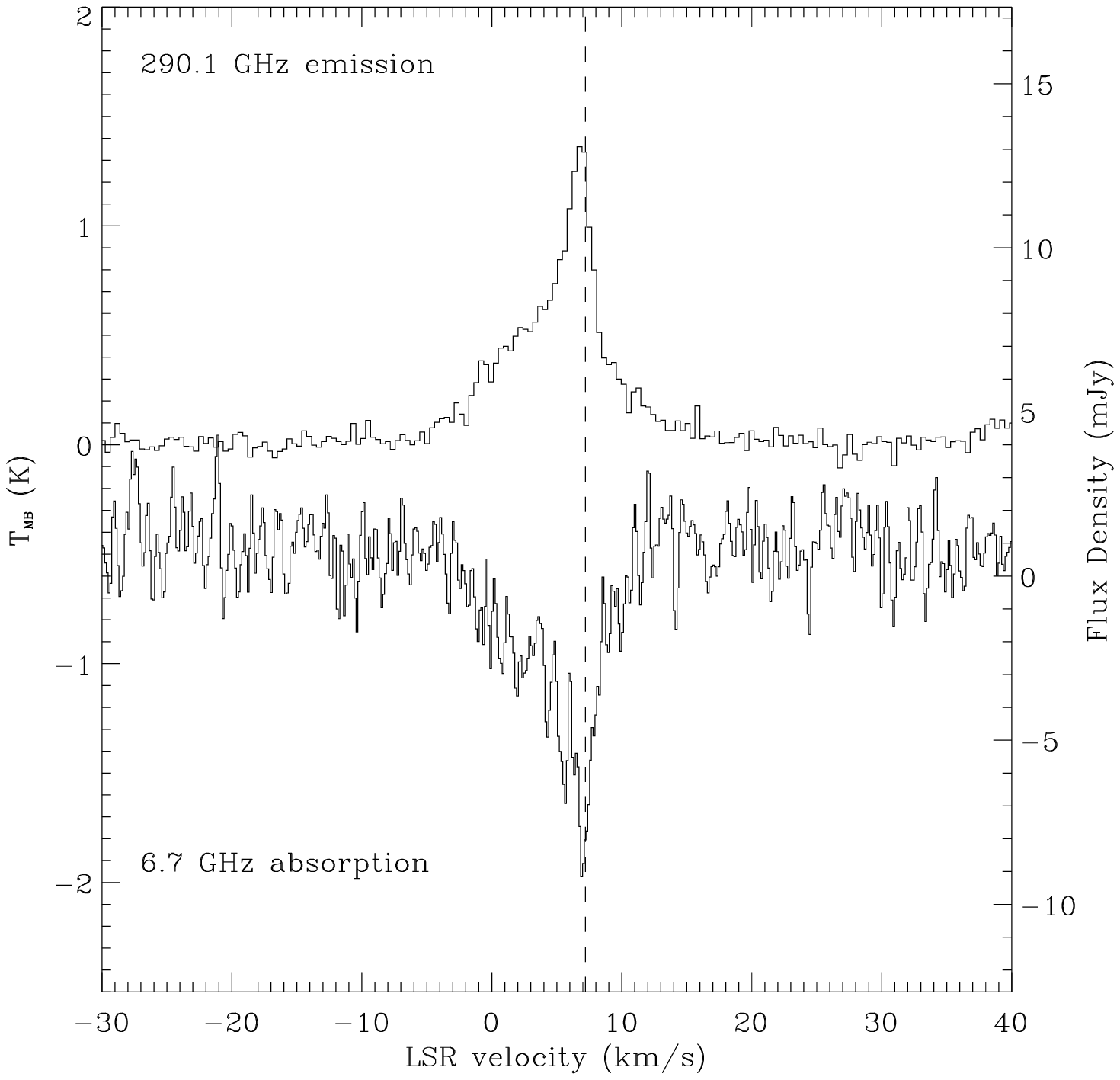}
\caption{A comparison of the line profiles of 290.1 GHz emission (left axis) and 6.7 GHz absorption (right axis) of methanol in NGC1333-IRAS~4A. The location of peak emission/absorption is indicated by the dashed line. The similarity of the line profiles indicates that the 6.7 GHz line is being absorbed over the full spatial extent of thermal methanol emission.}\label{iras4aprofile}
\end{figure}

\begin{figure*}
\centerline{\resizebox{1.0\hsize}{!}{\includegraphics[angle=270]{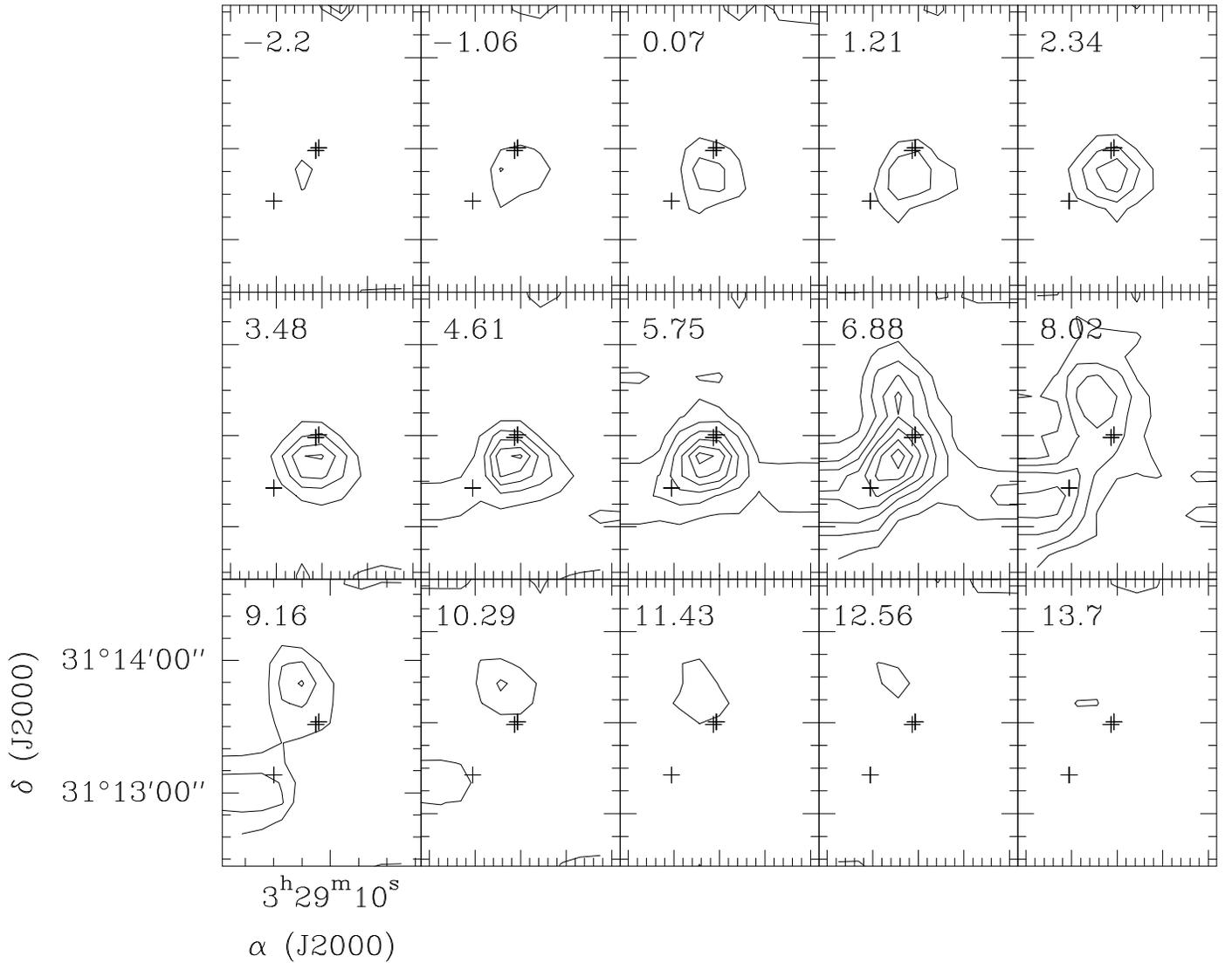}}}
\caption{Channel maps of the CH$_3$OH $6_0-5_0$ transition at 290.11 GHz
obtained with APEX and smoothed to a spectral resolution of 1.1 km~s$^{-1}$. The channel velocity is written in the top left corner of each panel. The crosses mark the positions of the protostars IRAS~4A1, 4A2, and 4B \citep{loon00}. The emission below 4~km~s$^{-1}$ and above 11~km~s$^{-1}$ is associated with the bipolar outflow powered by IRAS~4A. The contour levels are at integer multiples of 0.25 K.}\label{methanolchannelmap}
\end{figure*}

Fig. \ref{iras4aprofile} shows the 6.7 GHz absorption spectrum
measured towards IRAS~4A and the 290.1 GHz thermal methanol emission over the
same velocity range. The similarity of the line shapes of the two
lines is striking. From the channel maps of the 290.1 GHz line
(Fig. \ref{methanolchannelmap}), it can be seen that the emission
around 0 \kms~and 11 \kms~(LSR) arises from the blue and red-shifted
lobes respectively of the outflow in IRAS~4A. In contrast, the
continuum emission from the region at centimeter wavelengths is
concentrated over a $0.42\arcsec \times 0.28\arcsec$ region at the
core \citep{reip02}. Moreover, the continuum flux density at 3.6 cm is 0.43 mJy
(adding the emission from IRAS~4A1 and IRAS~4A2 components;
\citealt{reip02}) which is more than an order of magnitude weaker than
the observed depth of absorption (several orders of magnitude weaker when accounting for beam dilution). This indicates that the 6.7 GHz line
is being absorbed against the cosmic microwave background (CMB) over
the full spatial extent of thermal methanol emission (including the
outflow), as opposed to absorption against the continuum from the
core. This is similar to the observation of 12.2 GHz absorption
against the CMB in dark clouds by \citet{walm88}.

\section{Statistical equilibrium calculations}
\subsection{Anti-inversion in the $5_1-6_0~A^+$ 6.7~GHz transition}
As first realized by \citet{batr87}, methanol masers come in two varieties, termed Class I and II by \citet{ment91a,ment91b}. Class II methanol masers are closely associated with young high-mass (proto)stellar objects and arise from the same regions as hydroxyl (OH) masers. Like OH masers, Class II methanol masers are most likely pumped by far-infrared radiation in dense ($n \sim 10^7$~cm$^{-3}$), warm ($T \sim$ 150 K) gas \citep{cesa91,crag02,sobo94b,sobo94a,sobo97}.
In contrast, Class I methanol masers are frequently found in the general vicinity of intermediate and high-mass star formation, but often significantly offset (up to a parsec) from prominent center of activity, such as infrared sources or compact radio sources \citep[e.g., ][]{ment86}. Moreover, their pumping mechanism can be explained from basic properties of the methanol molecule \citep{lees73,lees1974}.  Often, Class I methanol masers are accompanied by absorption in the Class II maser lines against either the CMB \citep[e.g.,][]{walm88,whit89} or the continuum of the source \citep{whit88,ment91b,peng92}. The excitation properties of the 12.2~GHz line in absence of a strong infrared radiation field were explained in detail by \citet{walm88}. The 6.7 GHz line was predicted to be anti-inverted (i.e. over-cooled) under conditions similar to those producing anti-inversion in the 12.2 GHz line by \citet{ment91b} and \citet{crag92}. To our knowledge, the 6.7 GHz line has been  detected in absorption mostly against the radio continuum of the source \citep[e.g., ][]{ment91b,impe08}, and the only detections of absorption against the CMB are what is reported here, and probably, in the star forming region NGC~2264 \citep{ment91b}.

To investigate the physical conditions leading to absorption of the 6.7 GHz line against the CMB, we carried out a large velocity gradient (LVG) analysis with spherical geometry \citep{leur04,leur07}. In contrast to the approach used by  \citet{leur07} for modeling high-mass star forming regions, the only external radiation field used in our calculations is that of the CMB. Moreover, only a sub-sample of energy levels (the first 100 levels) was used for the calculations.

\begin{figure*}
\centering
\subfigure[]{\includegraphics[width=7cm]{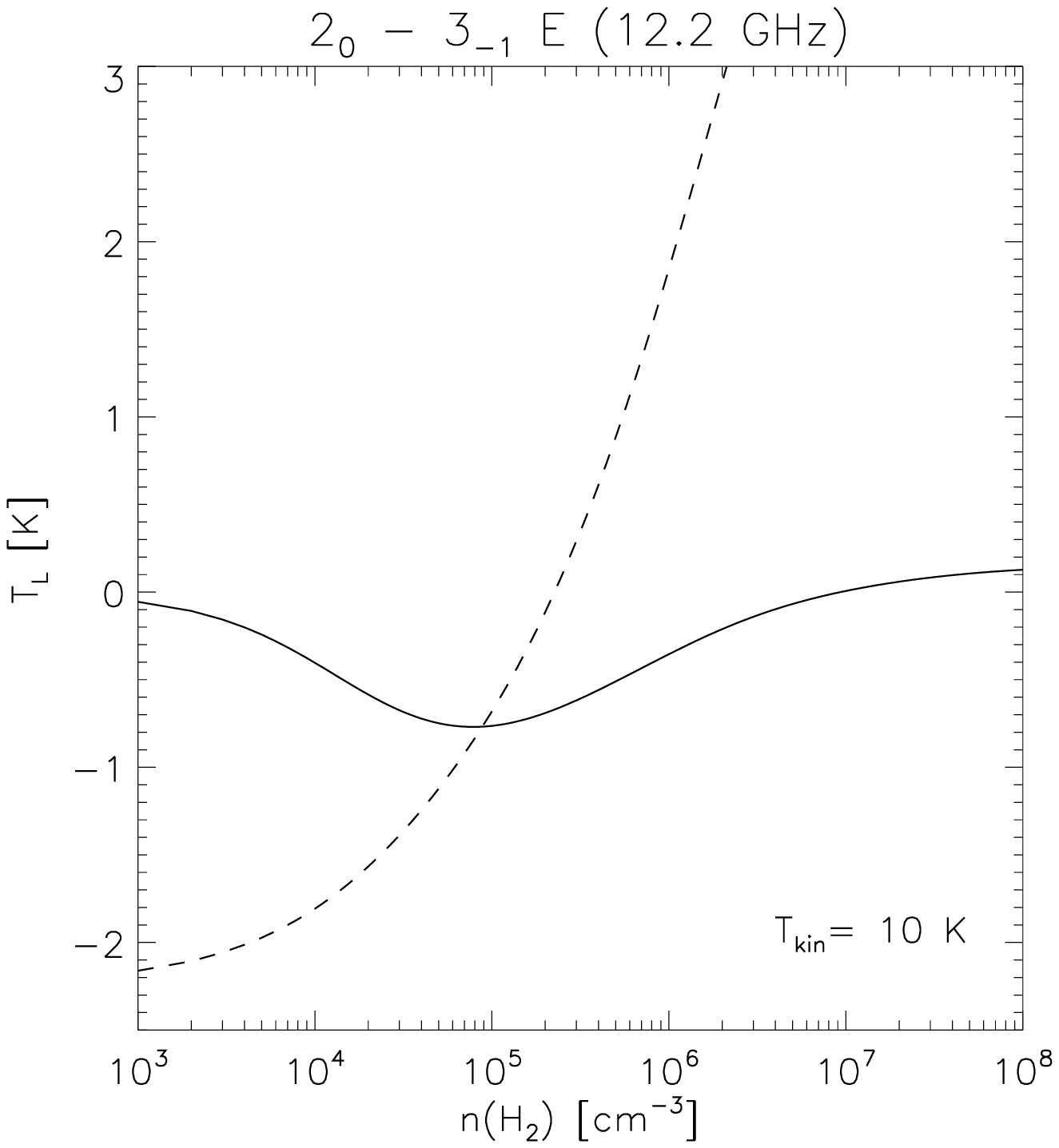}}
\subfigure[]{\includegraphics[width=7cm]{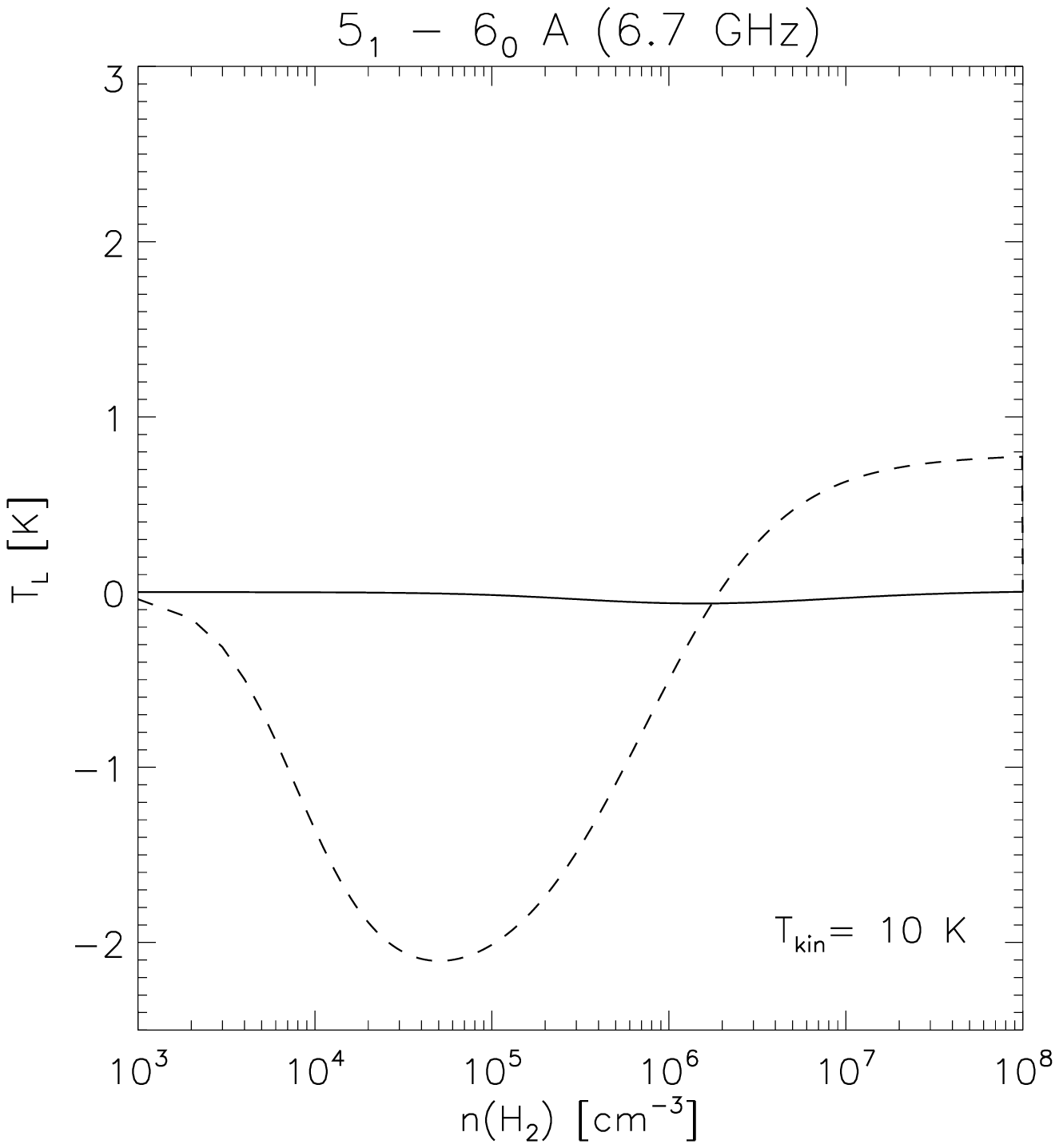}}
\subfigure[]{\includegraphics[width=7cm]{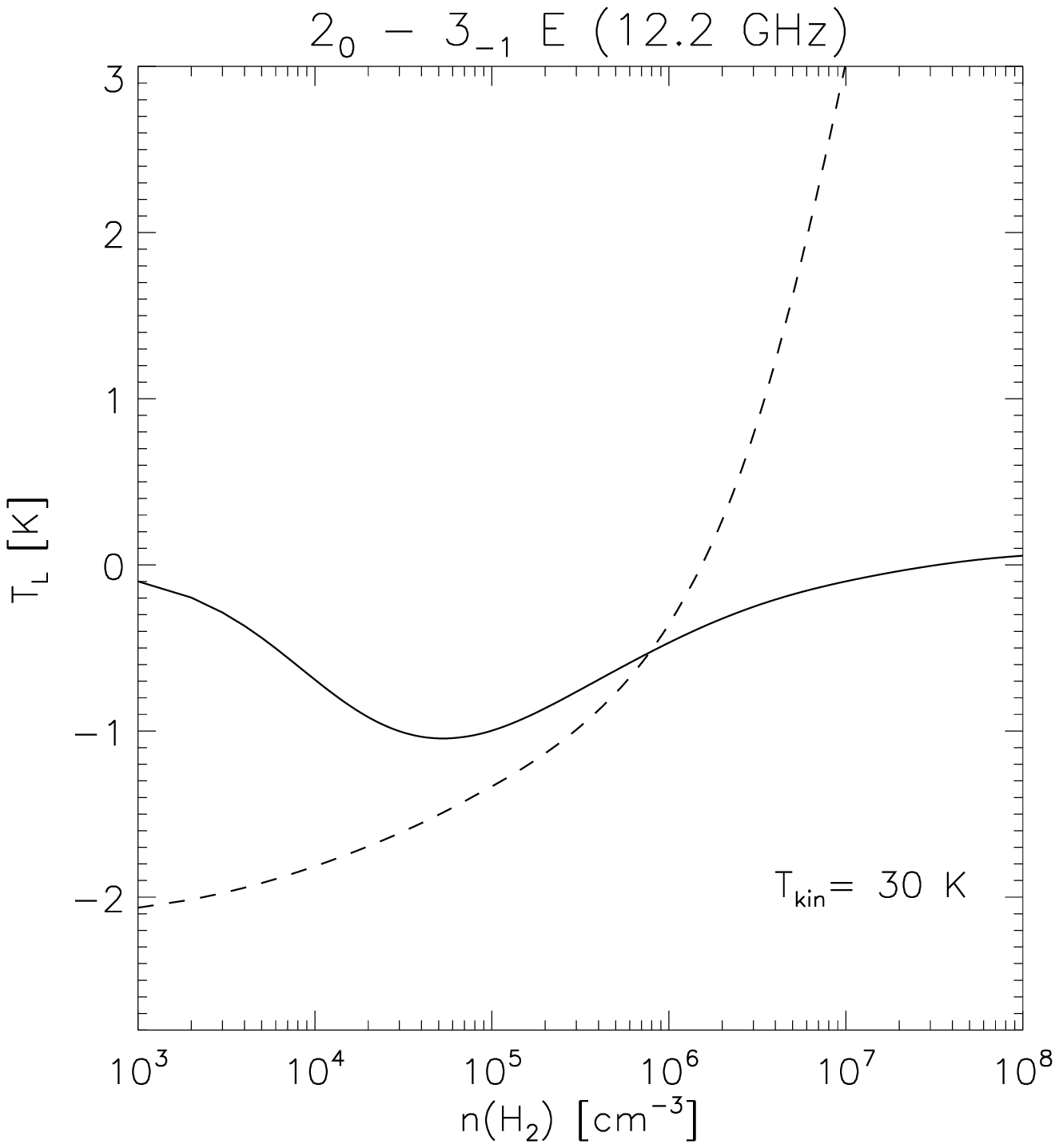}}
\subfigure[]{\includegraphics[width=7cm]{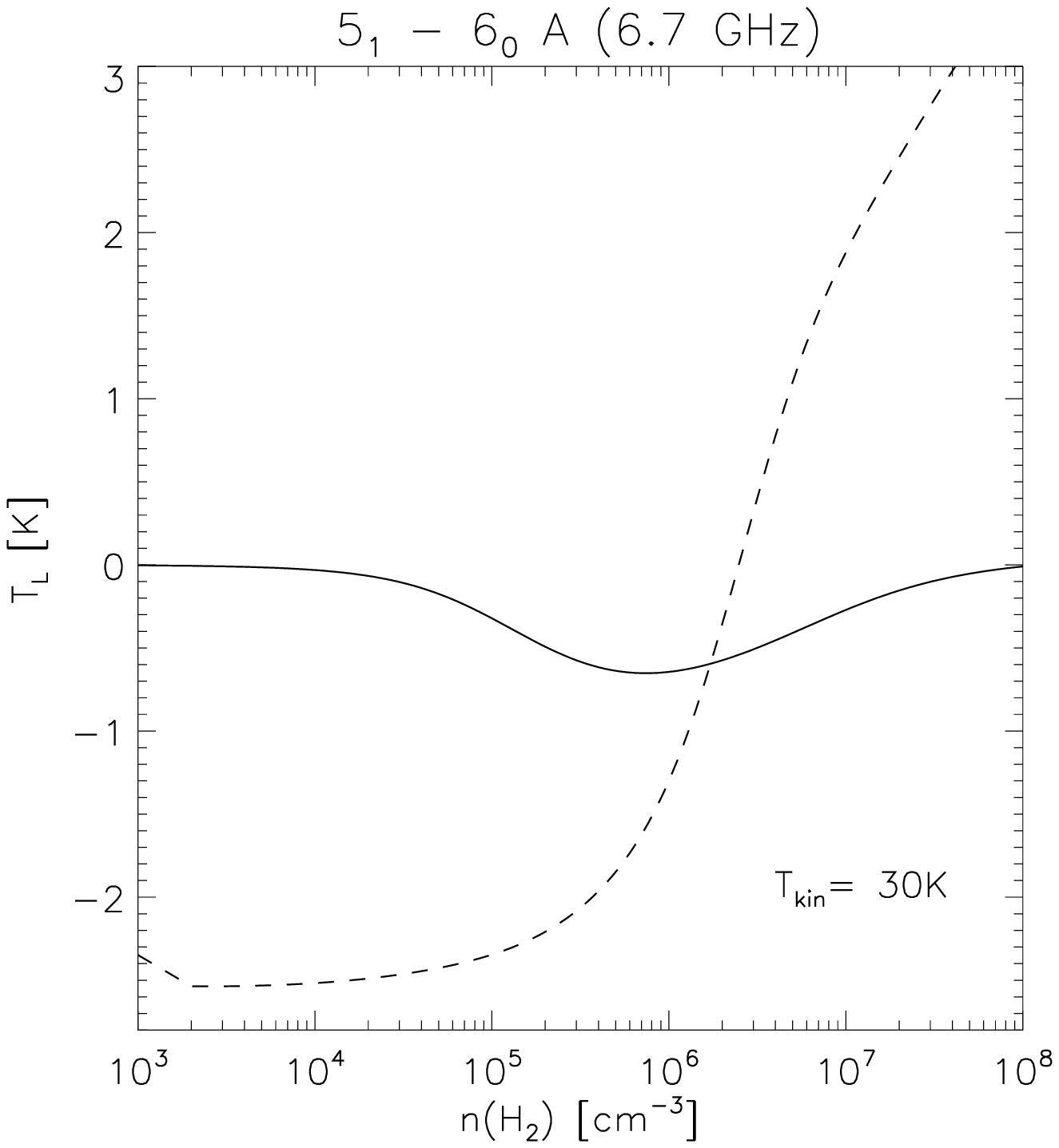}}
\subfigure[]{\includegraphics[width=7cm]{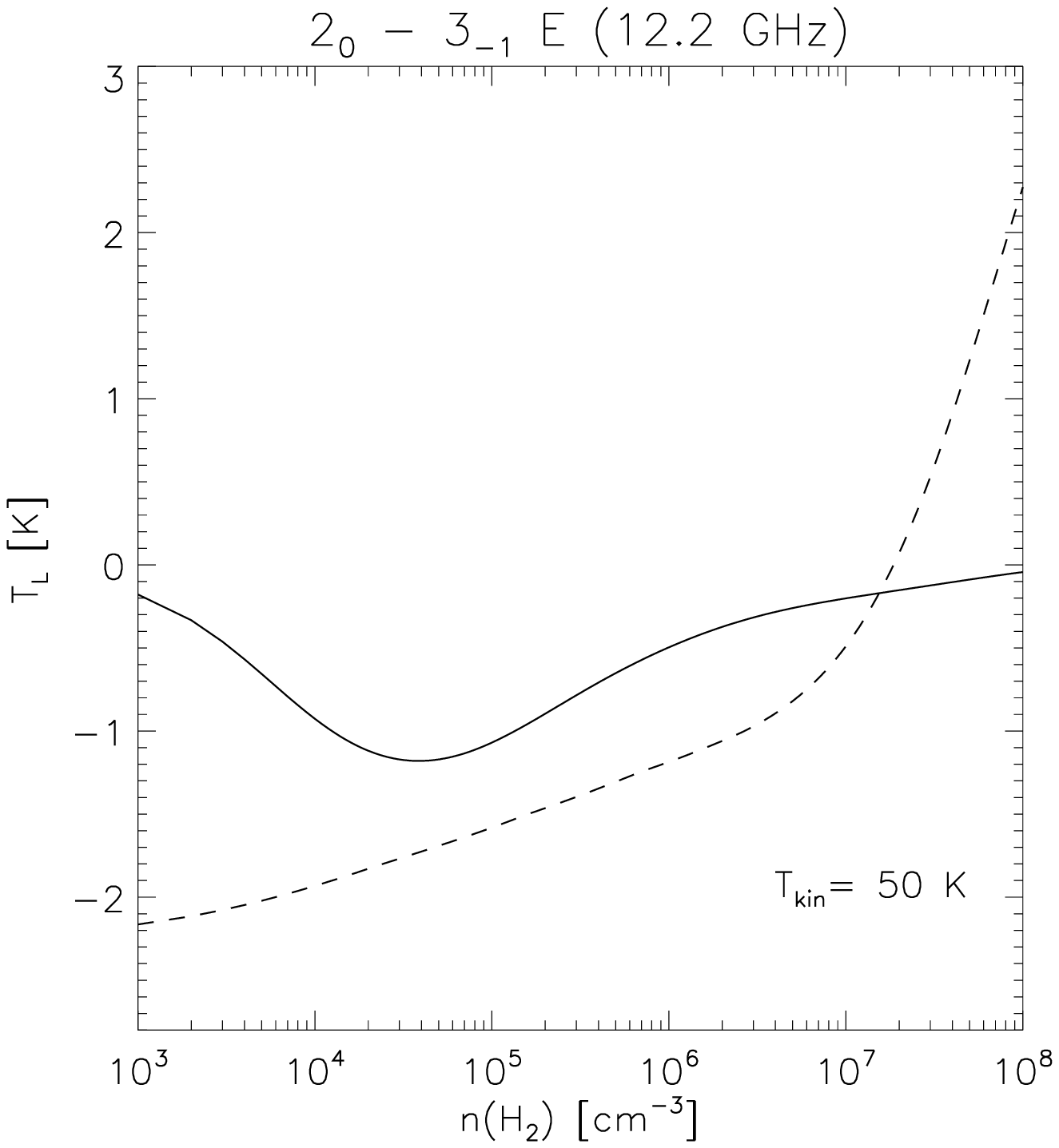}}
\subfigure[]{\includegraphics[width=7cm]{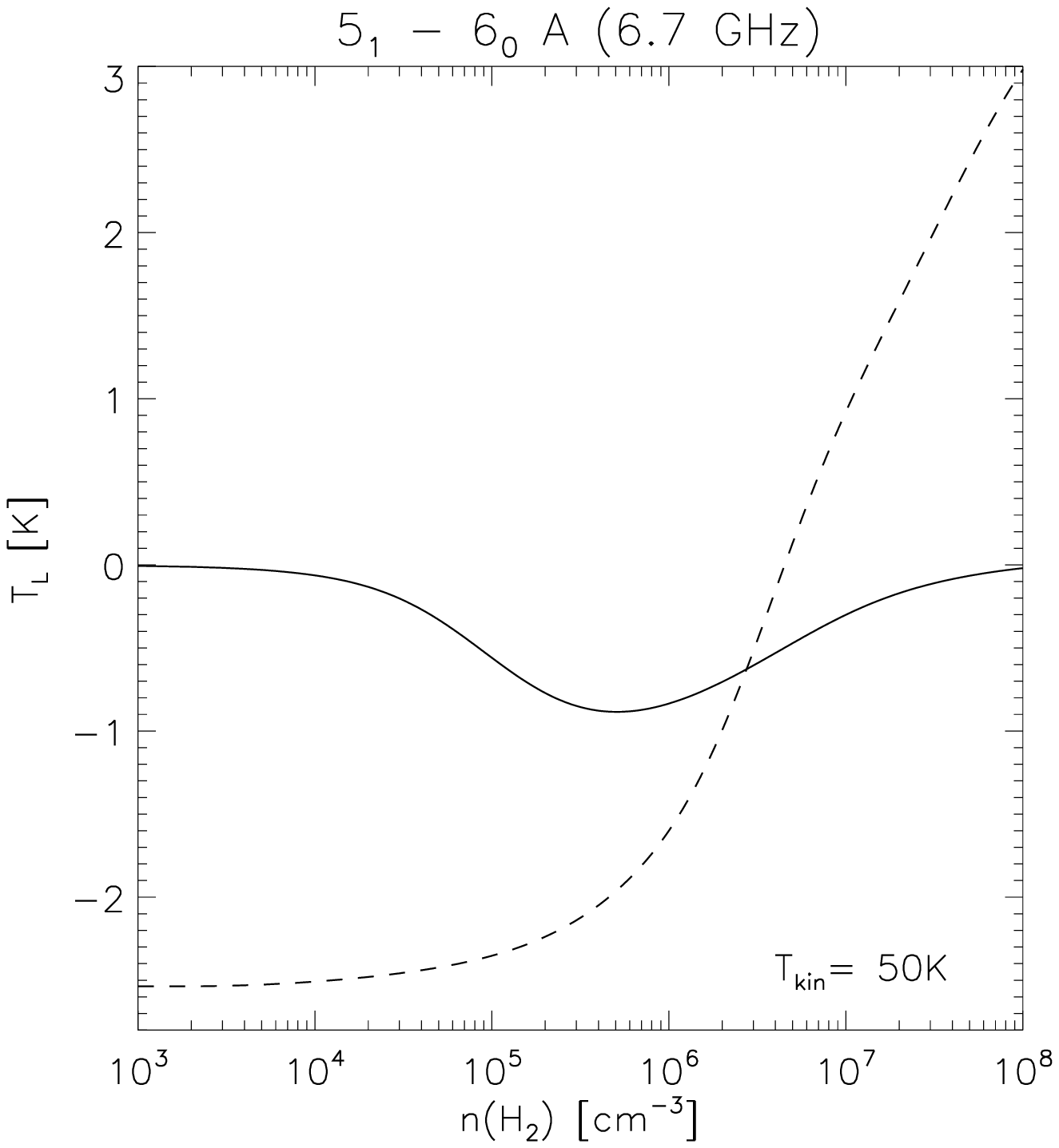}}

\caption{Results of statistical equilibrium calculations for the
$2_0-3_{-1}$-$E$ (left column) and $5_1-6_0$-$A^+$ (right column) transitions. The solid and dashed lines show the predicted line intensities for CH$_3$OH column densities (for both $E$ and $A$ species) of 10$^{14}$~cm$^{-2}$ and 10$^{16}$~cm$^{-2}$ respectively, and a linewidth of 1 \kms.\label{6GHz_12GHz}}
\end{figure*}

Figure~\ref{6GHz_12GHz} shows the results of our calculations for the $5_1-6_0~A^+$ transition at 6.7~GHz, for two different CH$_3$OH-$A$ column densities ($10^{14}$ and $10^{16}$~cm$^{-2}$), at temperatures of 10, 30 and 50~K. A linewidth of 1 \kms~was assumed for all cases. For comparison, the $2_0-3_{-1}$-$E$ line at 12.2~GHz is also shown. It can be seen that both lines have a very similar behavior. They are strongly anti-inverted at low densities ($n \le 10^6$~cm$^{-3}$) in absence of an infrared radiation field, the primary difference being that the 6.7~GHz line is quenched at higher densities compared to the 12.2~GHz line. For both lines, the intensity of absorption increases with abundance of CH$_3$OH. Moreover, the density at which the absorption line intensity is maximum, decreases with increasing temperature. Since the only radiation field in our calculations is that of the CMB, our results are not applicable for sources that have an IR radiation field, whose effect among others is to produce strong masers in both lines. For a discussion of the excitation of the 6.7 and 12.2~GHz in the presence of IR radiation fields, we refer to \citet{sobo94a} and \citet{sobo97}, or to the more recent work of \citet{crag05}.

We can verify the predictions of our analysis by simultaneously modeling the CH$_3$OH spectrum at 6.7~GHz and 290~GHz, assuming that the lines at the two wavelengths trace  the same gas. In the next section, we will model the CH$_3$OH emission of NGC1333-IRAS~4A.  As discussed in section \S \ref{sec_res}, we refrain from a similar analysis for IRAS~4B, as the source lies near the edge of the area mapped with APEX, preventing us from obtaining a spectrum of the source at the spatial resolution of the Arecibo data.

\subsection{Methanol around IRAS 4A}
As explained in \S \ref{sec_res}, to model the two datasets, we smoothed
the 290~GHz spectrum to the spatial resolution of the 6.7~GHz observations (HPBW of 40$''$). The similarity of the line profiles at 6.7~GHz and 290~GHz (Fig.~\ref{iras4aprofile}) strongly suggests that the absorption at 6.7~GHz arises from the same gas that gives rise to 290~GHz emission. Hence, to interpret the APEX and Arecibo CH$_3$OH data, we applied the method described by \citet{leur04}, based on an LVG analysis and simultaneous fitting of the entire spectrum with different non-interacting velocity components, each of them being characterized by its source size, kinetic temperature, column density and density.

The interpretation of our observations is rendered difficult by the
complexity of the IRAS~4 region and by the poor spatial resolution of our data. The $ 40''$ Arecibo beam includes
the hot corino IRAS~4A, and the molecular outflow originating from it
\citep{choi01,choi05}. As discussed in section \S \ref{sec_res},
contamination from IRAS~4B at this position is small enough to be neglected.

The line profiles are characterized by strong blue-shifted and weaker red-shifted emission. The same asymmetry between the blue- and red-shifted emission that can be seen in our CH$_3$OH data was observed by \citet{bell06} in CS.  From the channel maps at 21.7$''$ resolution of the CH$_3$OH $(6_K-5_K)$ band (Fig. \ref{methanolchannelmap}), we conclude that the blue-shifted and red-shifted emission come from compact sources to the north and south of IRAS~4A respectively. By fitting the peak of the integrated intensity map of the blue-shifted wing of the $6_0-5_0$-$A^+$ line at 290.1~GHz (in the range of --2.5 to 4~km~s$^{-1}$) with a two dimensional elliptical Gaussian and deconvolving for the APEX beam, we estimate a source size of 14$''$ for the blue lobe. A similar source size is obtained for the red lobe, using the integrated emission of the red-shifted wing of the H$_2$CO ($4_{0,4}-3_{0,3}$) line, which is stronger than the CH$_3$OH $(6_K-5_K)$ transitions. However, the signal to noise ratio in the red wing is lower than that of the blue wing, resulting in higher uncertainties.  Emission around the central velocity is centered close to IRAS~4A, but is elongated along the north-south direction and is probably contaminated by the blue and red lobes.  This is also seen in the CH$_3$OH ($6_{\pm 2}-5_{\pm 2}$)-$E$ lines, which are too weak for a channel map analysis, but whose integrated intensity map shows three peaks, corresponding to IRAS~4B, and the blue and red lobes of the molecular outflow from IRAS~4A. However, the peak corresponding to the blue lobe of the outflow extends to the core of IRAS~4A.

\begin{figure*}
\centering
\subfigure[]{\includegraphics[width=15cm]{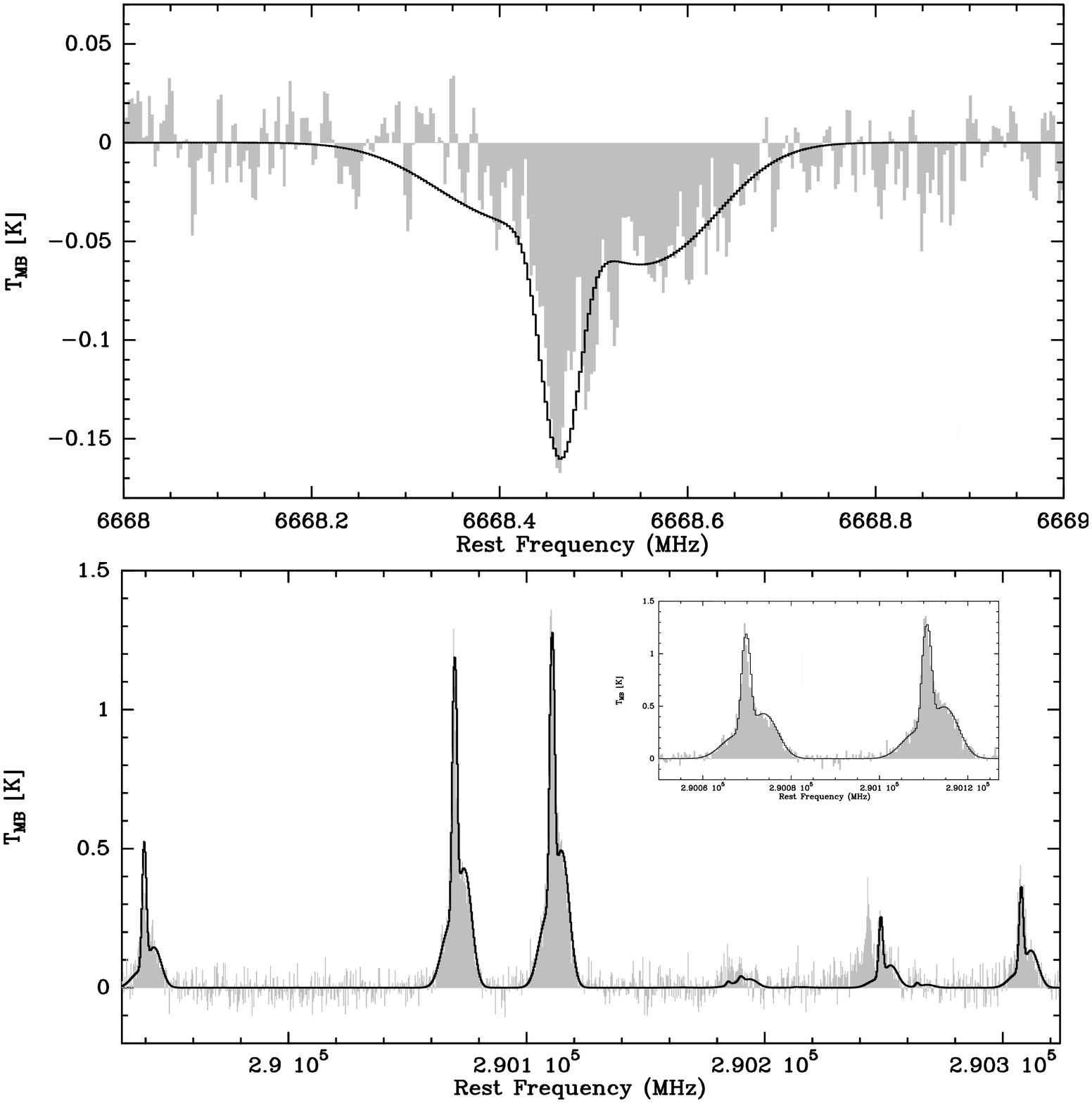}}
\caption{Methanol spectra at 6.7~GHz (top) and 290~GHz (bottom) towards NGC1333-IRAS~4A. The solid line represents the synthetic spectrum for the best fit parameters. The inset in the bottom panel shows the $6_{-1}-5_{-1}$-$E$ and $6_0-5_0$-$A^+$ lines in greater detail.\label{model}}
\end{figure*}

We modeled the spectrum using three velocity components, each having a compact size of 14$''$, corresponding to the blue- and red- shifted emission and
the hot corino in IRAS~4A. Since the red lobe emission is detected in fewer transitions, we mostly use this component more to obtain a reasonable fit of the line profiles than to derive its physical conditions. Figure~\ref{model} shows the best fit model spectrum for the 6.7.~GHz line and the $(6_K-5_K)$ band, overlaid on the observed spectra. The best fit parameters are:
$T=17$~K, $n=2~10^6$~cm$^{-3}$, and $N(\rm{CH_3OH})=4~10^{15}$~cm$^{-2}$ for the core velocity component;
$T=27$~K, $n=4~10^6$~cm$^{-3}$, $N(\rm{CH_3OH})=2~10^{15}$~cm$^{-2}$ for the blue-shifted component; and
$T=29$~K, $n=1~10^6$~cm$^{-3}$, $N(\rm{CH_3OH})=10^{15}$~cm$^{-2}$ for the red-shifted component.
The linewidths used in the model are 1.9, 7 and 8 \kms~for the core, blue-shifted and red-shifted components, respectively.The total column densities are derived assuming the same abundance for the two symmetry states, $A$ and $E$.

The 290~GHz observations include the $3_0-2_{-1}~E$ line of methanol at 302.369 GHz from the upper sideband of the APEX-2A receiver. However, since the sideband ratio of the receiver as a function of frequency is not well known, we did not include this line in our calculations. We did compare the line intensity predicted by the best fit model with the intensity in our data. The model predicts a line intensity $T_{MB}\sim 0.6$~K against the observed value of 0.2 K. Since the sideband ratio of the receiver at this frequency is not known, and since no other solution that fit all lines was found, we believe our results to be compatible with the data.

We refrain from a more rigorous analysis of our results due to the complexity of the region, coarse resolution of the data, and the limited number of transitions available for modeling. Our observations are mostly sensitive to the density of the gas and the CH$_3$OH column density. However, since the emission is optically thin, similar results can be achieved by increasing the column density of one of the components and decreasing its corresponding source size and vice versa. Finally, the low temperature derived by our analysis is mostly constrained by the weakness of the 6.7~GHz absorption. Although the fit is not excellent (with a reduced $\chi^2$ of 2), the model predictions for the 6.7~GHz absorption are compatible with our observations at 290~GHz.

\subsection{Why do we not observe class II methanol maser emission?}
According to \citet{crag05} the 6.7 GHz CH$_3$OH line shows strong maser action for dust temperatures, $T_D$, above 120 K. To estimate the radius, $r_{120}$, of the region around IRAS 4A within which that value is attained, we use the 
Stefan-Boltzmann law, since the dust emission is certainly optically thick at all relevant wavelengths.
We assume a bolometric luminosity, $L_*$, of $11.6 L\odot$, where we have scaled the estimate of \citet{sandell1991} to the Hipparcos distance of the Per OB2 association, i.e., 318 pc \citep{dezeeuw1999}. We calculate $r_{120}$ to be 37 AU. Optically thin dust emission, e.g. with an emissivity index, $\beta$, of 0.85, derived by \citet[see their eq. (1)]{bell06}, would yield an even smaller $r_{120}$ of 22 AU.  Modeling IRAS 4A's circumstellar environment as a collapsing envelope, \citet{bell06} derive relations between the density, $n$, and radial distance, $r$, for $r > 500$ AU 
at different times. At $r = 500$ AU, their ``best fit'' model predicts a density of $n = 10^7$~cm$^{-3}$ and their data suggests a density that may even be a factor of 3 greater.  While extrapolating the density to $r_{120}$ (37 AU) may not be straightforward, it is almost certain that the density that close to the central object will be vastly in excess of the $10^8$~cm$^{-3}$ at which any maser action would be quenched \citep[see Fig. 3 of ][]{crag05}. 

This realization may be extended to low-mass protostars/hot corinos in general and explains the notorious absence of class II methanol masers around such objects \citep[see, a.o,  the extensive survey of ][]{mini03}. In plain words: for low-mass protostellar objects, the region in which the radiation field would allow maser pumping is so dense that the maser emission is quenched, i.e., the 6.7 GHz transition's energy levels are thermalized.

\section{Conclusions}
We have made the first conclusive detection of absorption of the 6.7 GHz line against the CMB towards hot corino sources NGC1333-IRAS~4A and NGC1333-IRAS~4B. A LVG analysis indicates that the 6.7 GHz line is indeed strongly anti-inverted for densities lower than $10^6$ \cmiii~almost independent of temperature. Using observations of submillimeter $6_K-5_K$ lines of methanol using the APEX telescope, we model NGC1333-IRAS~4A, and are able to reproduce the observed 6.7 GHz absorption using cold ($T\sim 15-30$ K), dense (n $\sim 10^6$ \cmiii) gas. The absence of maser emission at 6.7 GHz in hot corinos, and low-mass protostellar objects in general, is presumably due to the very high densities in the region where the radiation field would provide for maser pumping.

\begin{acknowledgements}
This work was supported in part by the Jet Propulsion Laboratory, California Institute of Technology, under a contract with the National Aeronautics and Space Administration. This research has made use of NASA's Astrophysics Data System.
\end{acknowledgements}

\bibliographystyle{aa}

\bibliography{0556references}

\end{document}